# Mono-energetic proton beams accelerated by a circularly polarized laser pulse with a thin solid foil

X.Q. Yan[1*], C. Lin[1], Z.M. Sheng[1,2], Z.Y. Guo[1], B.C. Liu[3,1], Y.R. Lu[1], J.X. Fang[1], J.E. Chen[1]

[1] *State Key Laboratory of Nuclear Physics and Technology, Institute of Heavy Ion Physics, Peking University, Beijing 100871, China*
[2] *Laboratory of Optical Physics, Institute of Physics, Chinese Academy of Science, Beijing 100080, China*
[3] *Graduate University, Chinese Academy of Science, Beijing 100080, China*

**Abstract**

The acceleration of ions in the interaction of a circular polarized laser pulse with a thin foil is investigated. For circular polarization laser pulses, the quasi-equilibrium for electrons is established due to the light pressure and the electrostatic field built up at the interacting front of the laser pulse. The ions located within the skin-depth of the laser pulse can be synchronously accelerated and bunched in the charge couple processes by the electrostatic field, and thereby a monoenergetic and high intensity ion beam can be generated, which has been proved by 1D and 2D particle-in-cell simulations.

---
[*] Email: X.Yan@pku.edu.cn

State-of-the-art lasers can deliver ultraintense, ultrashort laser pulses, with intensities exceeding $10^{21}$W/cm$^2$, with very high contrast ratios, in excess of $10^{10}$:1. These systems can avoid the formation of plasma by the prepulse, thus opening the way to laser-solid interactions with ultra-thin solid targets [1]. Solid targets irradiated by a short pulse laser can be an efficient and flexible source of MeV protons as well as highly charged MeV ions. Such proton beams are already applied to produce high-energy density matter [2] or to radiograph transient processes [3], and they offer promising prospects for tumor therapy [4], isotope generation for positron emission tomography [5], and fast ignition of fusion cores [6].

In the intense-laser interaction with solid foils, usually there are three groups of accelerated ions. The first two occur at the front surface, moving backward and forward, respectively [7,8,9], and the third one occurs at the rear surface (TNSA) [10,11]. In these acceleration methods, usually the linear polarized (LP) laser pulse is used to heat the electrons by the J×B heating[12]. However, energetic electrons fly quickly away and the ions couldn't catch them, so a significant fraction of the absorbed laser energy is contained in the electrons and the efficiency of laser energy transformation to ions is small (less than 1%). Fast electrons may also cause a premature explosion of the thin foil. Furthermore, as the output beams in these methods are accelerated only by electrostatic fields and have no longitudinal bunching in (x, $p_x$) plane, their distribution profiles are exponential with nearly 100% energy spread. Simulations and experiments have shown improved efficiency and higher ion energy with the use of thinner targets, enhanced TNSA (BOA) [13] method is proposed to improve the energy efficiency. However the ions beam intensity is not improved, because there is no longitudinal bunching process. This paper proposes a new acceleration mechanism, where the ion beams are generated from an ultrathin foil irradiated by an ultraintense laser pulse with the circular polarization (CP). The main difference is the utilization of a CP laser pulse that inhibits a strong electron heating. The high-density ion bunches can be generated in the interaction of a CP laser pulse with a dense target [14,15]. We will develop the idea in ref.[15] through a simple model and dynamic equations that enable us to describe the bunching and acceleration process .

At first we carried out numerical simulations by a fully relativistic one-dimensional (1D) PIC code [16]. In our simulation a CP laser pulse with normalized peak amplitude $a_L = 5$ and duration 100 $T_L$ is normally incident on a purely hydrogen plasma slab, where $T_L$ is a laser period. To simplify the model, at the first a purely hydrogen plasma slab is used and it has a step density profile with the initial normalized density $n_0/n_c = 10$ and thickness $D = 0.2\lambda_L$ ($\lambda_L$ is the laser wavelength in vacuum, $n_c$ is the critical density of the incident laser pulse). In simulations, the target boundary is located at $x = 10\lambda_L$ and the laser front impinges on it at $t = 10T_L$. We take 100 particles per cell per species and the cell size of $\lambda_L/100$.

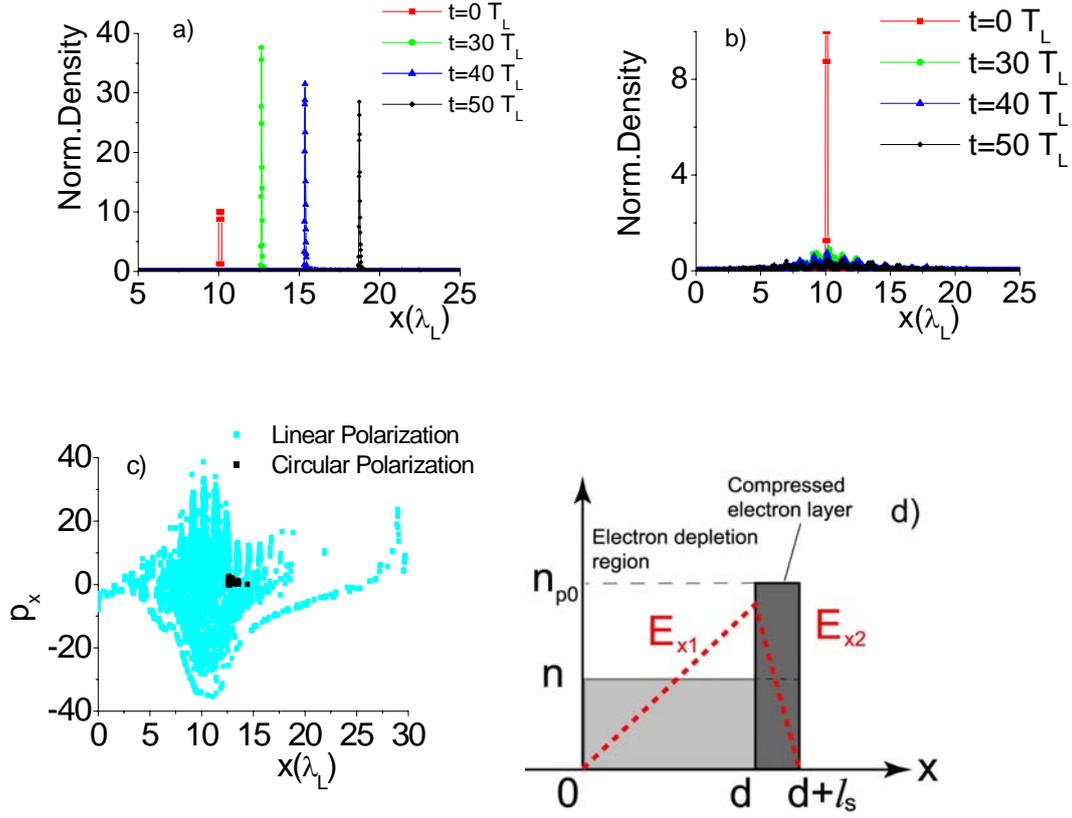

Fig.1 (color online) a) snapshots of electron density in CP case ; b) snapshots of electron density in LP;c) Phase space at t=30$T_L$; d) Schematic of the equilibrium density profiles of the ions n and the electron $n_{p0}$. The position at x=d indicates the electron front, there laser evanescence starts and vanishes at x=d+ $l_s$, where $l_s$ is the plasma skin depth.

Fig.1a shows the snapshots of electrons density when a CP laser pulse is normally incident on the target. Its oscillating part of the ponderomotive force is zero and the J×B heating [12] is less efficient (shown in Fig.1c), then electrons are mainly pushes forward and compressed into a narrow region. On the contrary, the electrons are obviously heated and fly away quickly in LP case, which causes the disassembly of the thin foil (see Fig.1b). In CP case nearly all electrons pile up in the front of laser pulse and make up of a compressed electron layer, leaving behind a charge depletion region and giving rise to a strong electrostatic field $E_x$ holding the layer back. The electron layer will be slowly accelerated and kept in a series of quasi-equilibriums between the electrostatic and the light pressure, this can be proved by the comparison between Fig.1c and Fig.4a. Therefore a **linear** analytical model can be built as Fig.1d shows. The induced electrostatic fields have linear profiles both in the depletion region ($E_{x1} = E_0 x/d$ for $0<x<d$) and in the electron layer ($E_{x2} = E_0(1-(x-d))/l_s$ for $d<x<d+l_s$). The parameters $E_0$, $n_{p0}$ and $l_s$ are related by equations $E_0 = 4\pi e n_0 D$ and $n_{p0} l_s = n(d+l_s) = n_0 D$. The ions located at initial positions $x<d$ are debunched (particle at the front gets higher energy gain and particle at the rear got lower energy gain, so their distance become more and more longer ). In the linear model, as the $E_{x1}$ increases with x, these ions will never reach those with x>d and be lost in the acceleration process (**see**

**Fig.3**). When the light pressure exerted on the electrons $(1+\eta)I_L/c$ (with $\eta$ is the reflecting efficiency) is assumed to be balanced by $E_0 e n_{p0} l_s / 2$, the electrostatic pressure from depletion region, the ions inside the compression layer ($d < x < d + l_s$) are bunched by the electrostatic field with the profile $E_{x2} = E_0(1-(x-d))/l_s$.

Assumed the compressed electron layer is kept in the equilibrium, ions inside this layer are pushed forward by $E_{x2}$. As long as these ions are accelerated, the electrons will follow them and catch ions until new balance is satisfied. In these **charge couple process**es, the equilibrium is transitorily lost and the electrons rearrange themselves quickly to provide a new equilibrium if the laser pulse is not over. Therefore the electrons stay in a series of quasi-equilibriums and the ions in the electron layer can be accelerated and bunched by the electrostatic field.

To describe the interaction between the ions and the electron layer, the dynamic equations for ions are necessary. We define $\xi = (x_i - x_r)$, $-l_s/2 \leq \xi \leq l_s/2$, where $x_r = d + l_s/2$ is the position of a reference particle (a special ion). All the ions inside the electron layer appear to oscillate around it. The force acting on a test ion is given by

$$F_i = q_i E_0 (1 - (x_i - d)/l_s) \tag{1}$$

Thus, the motion equation for the ion is:

$$\frac{d(m_i \gamma_i \dot{x}_i)}{dt} = \frac{q_i E_0}{m_i \gamma_i^3}(1-(x_i-d)/l_s) \quad \text{or} \quad \frac{d^2 x_i}{dt^2} = \frac{q_i E_0}{m_i \gamma_i^3}(1-(x_i-d)/l_s) \tag{2}$$

where $q_i$ and $m_i$ is the charge and mass of the ion, $\gamma_i$ is the relativistic factor for reference particle (ion). The phase motion ($\xi$, t) around the reference particle can be written as:

$$\ddot{\xi} = -\Omega^2 \xi, \quad \Omega^2 = \frac{q_i E_0}{m_i l_s \gamma_i^3} \tag{3}$$

If the $\gamma_i$ factor varies slowly and $E_0$ is assumed to be quasi-constant, then the longitudinal phase motion of ions ($\xi$, t) is a harmonic oscillation $\xi = \xi_0 \sin(\Omega t)$. Note that the harmonic oscillation amplitude determines the energy spread like $\Delta w / w_r \sim \xi_0 \Omega / p_r$, where $w_r$ and $p_r$ are the kinetic energy and momentum of the reference particle, respectively, and $\xi_0$ is the oscillating amplitude, which is given by $\xi_0 \leq l_s \sim c/\omega_p$. The harmonic oscillation frequency $\Omega$ decreases with the increase of particle energy. If we take the laser amplitude $a_L=5$, $n_0/n_c=10$, and $\gamma_i = 1$ for protons at the beginning, the first period of phase oscillation is about $8T_L$ and the period will be lengthened as long as $E_0$ and $1/\gamma_i$ are decreased. If the final energy of the reference particle $w_r$ = 300 MeV, the estimated energy spread $\Delta w/w_r$ will be less than 4%. These are verified in our

PIC simulation as shown later.

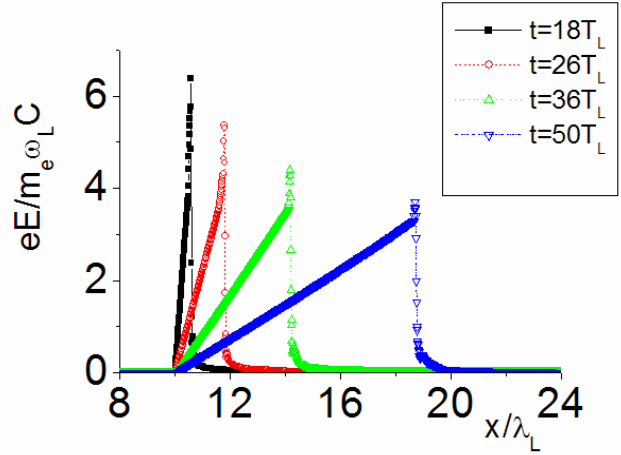

Fig. 2 (color online) Snapshots of the electrostatic field. The plasma slab has a step density profile with the initial normalized density $n_0/n_c = 10$ and thickness $D = 0.2\lambda_L$. The laser pulse has a normalized laser peak amplitude $a_L = 5$ and duration $100 T_L$.

Fig.2 shows the snapshots of the electrostatic field profile. Due to the ponderomotive force, an electron depletion region appears near the front. As the depletion region expands with time as the ion density is decreased, the electric field in this region extends over several wavelengths, so the slope of the field in the depletion region reduces gradually as shown in Fig. 2. In the compressed electron layer, it is found that the width of the layer remains to be equal to the skin depth ($l_s \cong \lambda_L/20$). Therefore the charge separation field in this layer nearly keeps the same steep linear profile, even though the maximum separation field is decreased slightly. It means the electrons and these ions are always confined near the compression region by each other. This is different from the surface acceleration with linearly-polarized light [7~10], where the ions see the accelerating field only in a short time. On the contrary, protons in the electron layer can be synchronously accelerated and bunched in this process, quite similar as in the radio frequency accelerator.

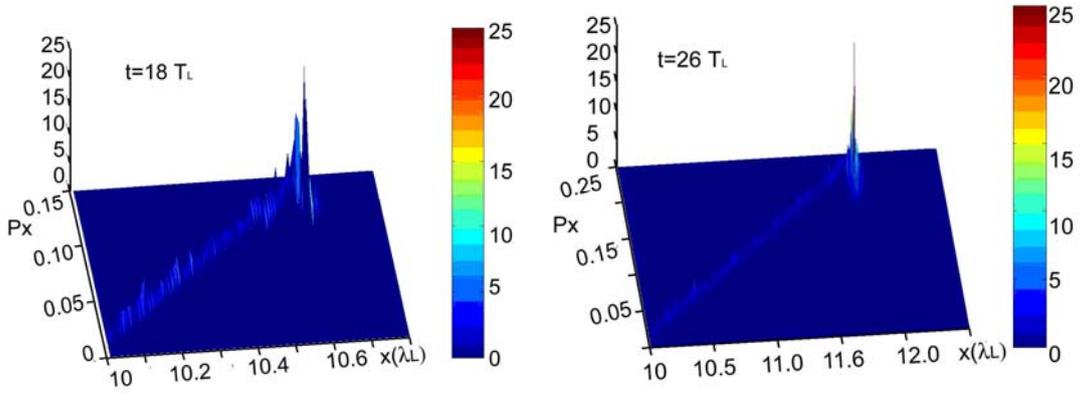

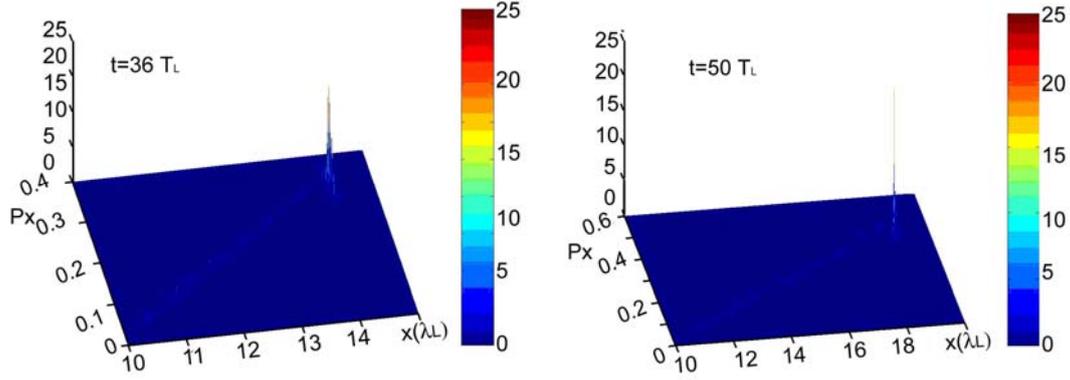

Fig. 3 (color online) Evolution of phase space distribution for ions, the 1st, 2nd, 3rd and 4th oscillation period are 8, 8, 10 and 14 $T_L$, respectively. The parameters for the laser pulse and plasma are the same as in Fig. 2. The longitudinal momenta of protons are normalized by $Mc$ with $M$ the rest mass of protons.

Fig.3 plots snapshots of the phase space of the protons. It shows that protons execute periodic oscillations as described by Eq. (3). Note that the laser pulse hits the plasma slab at $t=10T_L$. The initial oscillation period is just about $8T_L$, which is just equal to theoretical value estimated by Eq. (3). Later on, the oscillation period increases with time. **It shows that protons in the electron depletion region are debunched and the region is continuously elongated, while protons inside the electron layer are bunched and accelerated synchronously.** After about 40 laser cycles of the laser interaction, the protons have been accelerated to the longitudinal momenta about $p_x = 0.4Mc$, where $M$ is the rest mass of protons. The reflection coefficient is found to be about $\eta = 0.38$, which implies that more than 60% of the laser energy is converted into the electron and proton energy. The extremely high conversion efficiency results from the dynamic interaction between electrons and ions like in the laser-piston regime [17]. Among the absorbed laser energy, more than 99% is contained in the accelerated protons.

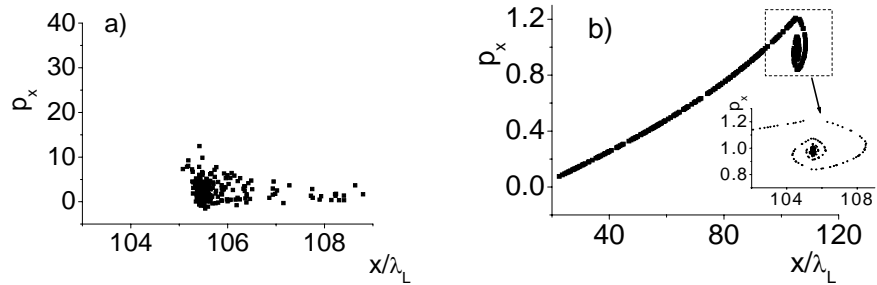

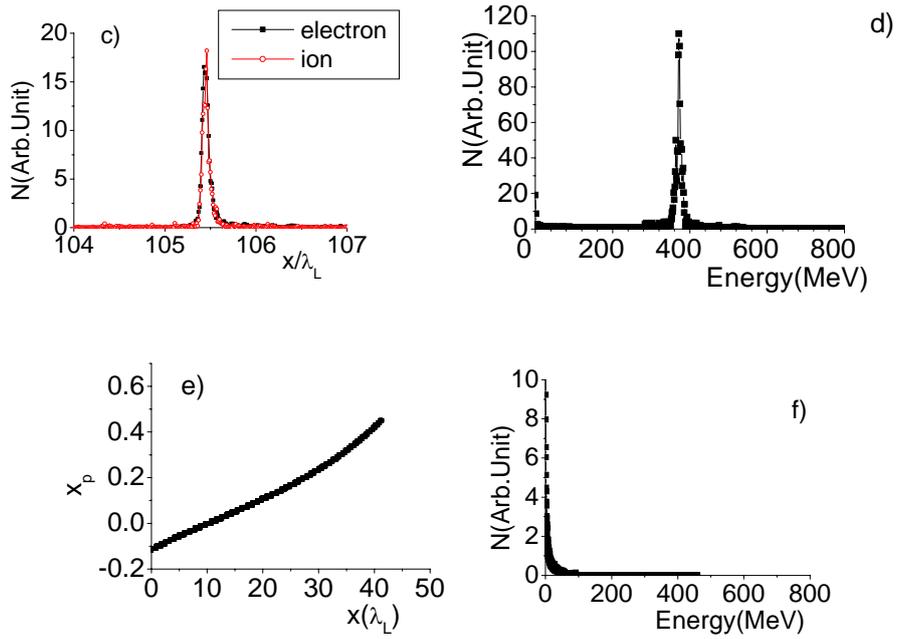

Fig.4 (color online) a) Phase space distribution of electrons in CP case; b) Phase space distribution of protons in CP case;(c) Electrons and protons density profiles; d) Energy spectrum of protons in CP case ;e) Phase space of protons in LP case; f) Energy spectrum of protons in LP case. The results are found at t=200 $T_L$, when the laser pulse interaction is almost terminated. The laser and plasma parameters are the same as in Fig. 2.

The distributions of electrons and ions in the phase space are plotted in Fig.4a and Fig.4b. **Fig.4c shows the output beam is nearly charge-neutral, which is helpful to transport the beam in a long distance. The laser radiation pressure continues accelerate the neutral beam in these charge couple processes if the pulse is not over.** The debunched ions form a long tail in the phase space(see Fig.3), however, their density is many times more tenuous than in the electron layer, so they look disappearing in the energy spectrum as Fig.4b and Fig.4d show, this would explain why is really no broad background underneath the monoenergetic peak at 375MeV. The Fig.4e and Fig.4f give the phase space and the energy spectrum of protons driven by a LP laser pulse with the same parameters. Compared to Fig.4f, the proton beam driven by a CP laser not only has much more high energy, but also has an awfully lower FWHM energy spread (< 4%) and higher peak current. This energy spread is also completely in agreement with our earlier estimation based upon Eq. (3). Note that the proton bunch has an ultrashort length about the skin depth $l_s$ or about 250 attoseconds in time ($\lambda_L$=800 nm). The number of accelerated protons in the bunch is about $n_0 l_s \sigma$, where $\sigma$ is the focused beam spot area. This gives about $5 \times 10^{12}$ quasi-monoenergetic protons for a focused beam diameter of 40μm in the present simulation.

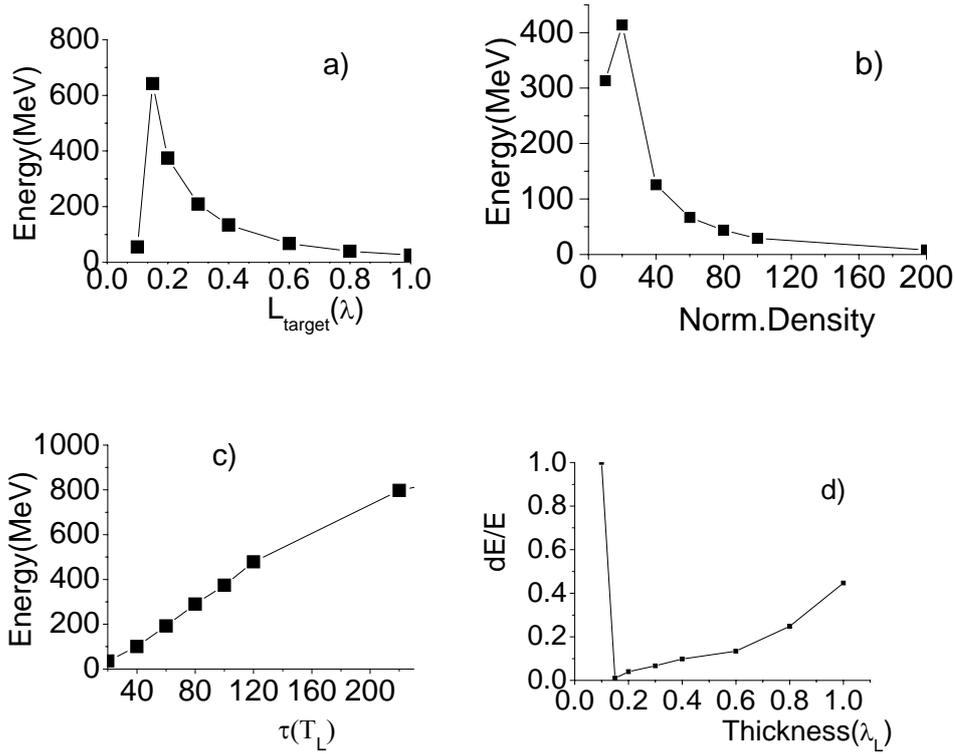

Fig. 5 a) Proton energy versus target thickness for the parameter of the normalized laser amplitude $a_L$=5, pulse duration $\tau=100T_L$, and the normalized plasma density $n_0/n_c=10$; b) Proton energy versus the plasma density with $a_L$=5, $\tau=100T_L$ and target thickness $D=0.1\lambda_L$; c) Proton energy versus laser pulse duration with $a_L$=5, the plasma density $n_0/n_c=10$, and target thickness $D=0.2\lambda_L$; d) Energy spread (FWHM) versus target thickness, $a_L$=5, $n_0/n_c=10$.

The proton energy is plotted versus the target thickness, the plasma density and the laser pulse length in Fig.5. The proton energy first increases with the target thickness. If the target is thinner than the skin depth of the laser pulse, the latter can transmit through the target, which results in a smaller radiation pressure. When the target thickness is close to the skin depth, only a small amount of the laser energy can transport through, so that ponderomotive force can induce a large electrostatic field for the proton acceleration. The maximum energy appears at about three times of the normal skin depth due to the relativistic effect [18] (see Fig.5a and Fig.5b). For an even thicker target or denser plasma, the highest energy of protons begins to decrease. Both Fig.5a and Fig.5b imply that mono-energetic proton beam can be generated, while Fig.5d shows the mono-energetic properties are affected by the target thickness. In order to get a higher energy and a lower energy spread, the plasma slab with three times skin depth is the best. Fig.5 c suggests that the proton energy increases almost linearly with the laser pulse duration for the longer acceleration time.

The results obtained so far are based upon the linear 1D model. We should take into account the numerous nonlinear and kinetic effects, as well as extend our considerations to a multidimensional geometry. The size of the simulation box is $100\lambda(x)\times 15\lambda(y)$. A circularly polarized laser pulse with dimensionless amplitude $a$ =5 propagates along the $x$ axis. The pulse size is $44\lambda\times 10\lambda$. The pulse has a trapezoidal shape (growth-plateau-decrease), with 2 λ−38 λ−2 λ in the $x$ direction, and

$2\lambda-6\lambda-2\lambda$ in the y directions. The target is a low density plastic foil(CH2) containing the carbon ions with the normalized density of 10 and protons with the density of 20. The plasma consists of three species: electrons, protons with $m_p/m_e$=1836, and carbon ions (with the charge $Z_i$ =+6) with $m_i/m_e$=1836×12.

The Fig.6a shows the electrons are compressed and pushed forward as a whole like in these 1D simulations, which is quite different from TNSA (BOA) mechanism [13]. The protons are bunched in the electron layer in the acceleration process (see Fig. 6c and Fig.6d), so a high current and monoenergetic proton beam is generated in Fig.6b. This shows that 2D effects do not substantially affect the acceleration process.

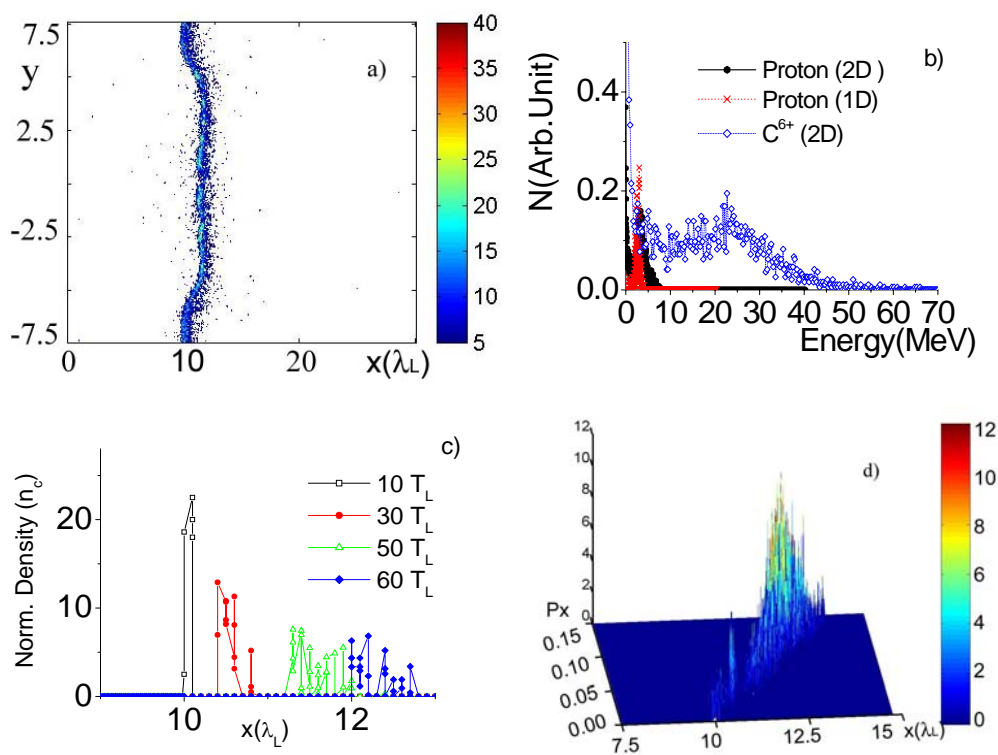

Fig.6 (color online)   a) Spatial distribution of electrons at t=60 $T_L$; b) Energy spectrum of protons and carbon (per nucleon) at t=60 $T_L$;   c) Evolution of the proton density in the target center; d) The phase space of protons at t=60 $T_L$. Circularly laser pulse with $a_L$=5, pulse duration $\tau$=42$T_L$, target thickness D= 0.1$\lambda_L$

In conclusion, a new mechanism of ions acceleration has been proposed, by which the ions can be synchronously accelerated and bunched by the electrostatic field like that in a radio frequency accelerator. With an ultrahigh intensity laser with the peak amplitude $a_L$=5 and pulse duration $\tau$=100$T_L$, one can obtain moderate monoenergetic ultrashort proton bunches with energy larger than 300MeV and an FWHM energy spread of only a few percent. Both 1D and 2D PIC simulations demonstrate that mono-energetic and high current proton beam can be generated by interaction of a circularly polarized laser pulse with a thin foil.

We would like to thank Dr.M.Chen from IOP of CAS for the discussions about the 2D PIC simulations. This work was supported in part by NSFC (under grants No. 10455001, 10605003,